\documentclass[english,aps, twocolumn, superscriptaddress]{revtex4-1}
\usepackage[T1]{fontenc}
\usepackage[latin9]{luainputenc}
\usepackage{textcomp}
\usepackage{amsbsy}
\usepackage{amstext}
\usepackage{graphicx}
\usepackage{esint}
\usepackage{babel}
\begin{document}

\title{Induction Mapping of the 3D-Modulated Spin Texture of Skyrmions in
Thin Helimagnets}

\author{S. Schneider}

\affiliation{Institute for Metallic Materials, IFW Dresden, Helmholtzstr. 20,
01069 Dresden, Germany}

\affiliation{Institut für Festkörper- und Materialphysik, Technische Universität
Dresden, 01062 Dresden, Germany}

\author{D. Wolf}

\affiliation{Institute for Solid State Research, IFW Dresden, Helmholtzstr. 20,
01069 Dresden, Germany}

\author{M. J. Stolt}

\affiliation{Department of Chemistry, University of Wisconsin-Madison, 1101 University
Avenue, Madison, Wisconsin 53706, United States}

\author{S. Jin}

\affiliation{Department of Chemistry, University of Wisconsin-Madison, 1101 University
Avenue, Madison, Wisconsin 53706, United States}

\author{D. Pohl}

\affiliation{Institute for Metallic Materials, IFW Dresden, Helmholtzstr. 20,
01069 Dresden, Germany}

\affiliation{Dresden Center for Nanoanalysis, Technische Universität Dresden,
01062 Dresden, Germany}

\author{B. Rellinghaus}

\affiliation{Institute for Metallic Materials, IFW Dresden, Helmholtzstr. 20,
01069 Dresden, Germany}

\affiliation{Dresden Center for Nanoanalysis, Technische Universität Dresden,
01062 Dresden, Germany}

\author{M. Schmidt }

\affiliation{Department Chemical Metal Science, Max Planck Institute for Chemical
Physics of Solids, Nöthnitzer Str. 40, 01187 Dresden, Germany}

\author{B. Büchner}

\affiliation{Institute for Solid State Research, IFW Dresden, Helmholtzstr. 20,
01069 Dresden, Germany}

\author{S. T. B. Goennenwein}

\affiliation{Institut für Festkörper- und Materialphysik, Technische Universität
Dresden, 01062 Dresden, Germany}

\affiliation{Center for Transport and Devices of Emergent Materials, Technische
Universität Dresden, 01062 Dresden, Germany}

\author{K. Nielsch}

\affiliation{Institute for Metallic Materials, IFW Dresden, Helmholtzstr. 20,
01069 Dresden, Germany}

\affiliation{Institute of Materials Science, Technische Universität Dresden, Helmholtzstr. 7,
01069 Dresden, Germany}

\author{A. Lubk}

\affiliation{Institute for Solid State Research, IFW Dresden, Helmholtzstr. 20,
01069 Dresden, Germany}
\begin{abstract}
Envisaged applications of skyrmions in magnetic memory and logic devices
crucially depend on the stability and mobility of these topologically
non-trivial magnetic textures in thin films. We present for the first
time quantitative maps of the magnetic induction that provide evidence
for a 3D modulation of the skyrmionic spin texture. The projected
in-plane magnetic induction maps as determined from in-line and off-axis
electron holography carry the clear signature of Bloch skyrmions.
However, the magnitude of this induction is much smaller than the
values expected for homogeneous Bloch skyrmions that extend throughout
the thickness of the film. This finding can only be understood, if
the underlying spin textures are modulated along the out-of-plane
$z$ direction. The projection of (the in-plane magnetic induction
of) helices is further found to exhibit thickness-dependent lateral
shifts, which show that this z modulation is accompanied by an (in-plane)
modulation along the $x$ and $y$ directions.
\end{abstract}
\maketitle

\section{Introduction}

Skyrmions \cite{Bogdanov1994} are topologically non-trivial vortex-like
spin textures, anticipated for application in spintronic technologies,
referred to as skyrmionics, in next generation magnetic data processing
and storage due to their facile manipulation by spin-polarized currents
of very low magnitude \cite{Nagaosa2013a,Kanazawa2017}. The unique
features of skyrmions, e.g., their dynamics, topological structure,
competing magnetic interactions, are generally of great interest from
a fundamental physics point, understanding emerging magnetic field-like
interactions induced by topologically non-trivial chiral spin structures.
In chiral-lattice ferromagnets without spatial inversion symmetry,
such as the B20 compound $\mathrm{Fe_{0.95}Co_{0.05}Ge}$ (see Fig.~\ref{fig: Fig. 2}~(a))
investigated in this work, skyrmions arise from the interplay between
the Dzyaloshinskii-Moriya interaction \cite{Dzyaloshinsky1958,Moriya1960}
and ferromagnetic exchange mechanisms \cite{Heisenberg1926}. Indeed,
these and similar competing interactions, such as surface dipolar
interaction, may lead to a whole zoo of non-trivial spin textures,
including helical, cycloidal and various skyrmionic phases (antiskyrmions
\cite{Nayak2017}, Néel skyrmions \cite{Kezsmarki2015}). Besides
spin-polarized STM \cite{Heinze2011a} and MFM \cite{Milde2013a}
probing the surface spin texture, X-ray microscopy \cite{Woo2016}
was used to investigate the projected magnetic structure of skyrmions.
Furthermore Lorentz transmission electron microscopy (TEM) and transport
of intensity (TIE) holography have been employed to reveal the projected
skyrmionic texture in a variety of studies \cite{Yu2010,Yu2011a,Seki2012c,Yu2013a}
in dependency of the applied magnetic field, temperature, sample thickness
and crystallographic orientation, covering a large class of materials.

However, in particular for skyrmionics, knowledge about the full three-dimensional
spin texture including its coupling to surfaces and interfaces, ubiquitous
in thin film technology, is of fundamental importance, because it
determines the stability and dynamics of the skyrmion state. Several
theoretic studies predicted the occurence of 3D modulated skyrmion
textures \cite{Rybakov2013,Leonov2016,Leonov2014,Rybakov2016} as
a consequence of surface anisotropies as well as 3D Dzyaloshinskii-Moriya
interaction, also taking into account similarities to smectic liquid
crystals \cite{Hinshaw1988,Glogarova1997}. In particular, Rybakov,
Borisov and Bogdanov theoretically predicted the presence of a chiral
surface twist \cite{Rybakov2013}, which was later experimentally
discovered in epilayers of chiral magnets \cite{Wilson2013a,Meynell2014a}.
This surface modulation renders the skyrmionic state stable in thin
film geometries (making it a ground state) as opposed to its metastable
(i.e., excited) nature in the bulk. Later, a full-blown phase diagram
of helical, skyrmionic, and other magnetic textures, such as Bobbers
\cite{Zheng2017} has been computed for thin films of isotropic chiral
magnets \cite{Rybakov2016}. Experimental studies on modulated 3D
spin textures in skyrmions have been, however, elusive to date. Similarly,
almost none of the abundant microscopy studies \cite{Yu2010,Yu2011a,Seki2012c,Yu2013a,Park2014,Kovacs2016a,Shibata2017,Jin2017}
give quantitative values of projected magnetic fields carrying a fingerprint
of the modulated 3D texture to the best knowledge of the authors.

Here, we seek to fill this gap by carrying out electron holography
(EH) studies at different orientations of the sample to quantitatively
reconstruct the projected magnetic field pertaining to both the helical
and the skyrmion lattice phase in single crystal nanoplates of the
isotropic chiral magnet $\mathrm{Fe_{0.95}Co_{0.05}Ge}$. We compare
our experimental results to magnetostatic simulations taking into
account 3D modulation models such the chiral surface twist in order
to discuss the presence of 3D spin textures in skyrmions. Our findings
clearly suggest (i) the presence of inhomogeneous spin textures similar
to previously discussed surface modulations and (ii) show that currently
available spin structure models cannot account for our experimental
results.

\section{Fundamentals}

Following \cite{Bogdanov1989} one may describe the isotropic chiral
magnet FeGe (space group $P2_{\text{1}}3$) with a continuum spin
model (normalized magnetization vector $\boldsymbol{m}$), i.e., as
a (meta)stable state of the free energy
\begin{eqnarray}
F & \propto & \int\bigg(J\left(\left|\nabla m_{x}\right|^{2}+\left|\nabla m_{y}\right|^{2}+\left|\nabla m_{z}\right|^{2}\right)+\label{eq: free energy}\\
 &  & D\boldsymbol{m}\left(\nabla\times\boldsymbol{m}\right)-\boldsymbol{H}\boldsymbol{m}M_{s}\bigg)\mathrm{d^{3}}r,\nonumber
\end{eqnarray}

the interplay of ferromagnetic exchange and the Dzyaloshinskii-Moriya
interaction leads to the formation of a helical spin order with a
periodicity $L_{D}=4\pi J/D$ determined by the ratio of the Dzyaloshinskii\textendash Moriya
interaction constant $D$ and the ferromagnetic exchange interaction
constant $J$ ($L_{D}\approx70\:nm$ in FeGe \cite{Lebech1999}).
When additionally applying a weak magnetic field below the critical
field $H_{D}=D^{2}/\left(2M_{S}J\right)$ skyrmions typically arrange
in a hexagonal lattice formed by three superimposing helical spin
waves appearing in a plane normal to the field irrespective of the
crystal orientation due to the weak crystal anisotropy in FeGe.

In addition, it is well-known that the stability of a skyrmion state
increases as the thickness of the FeGe sample decreases \cite{Yu2011a}.
Recently, a 3D modulation of the spin texture of the helical and Skyrmion
phase with a chiral surface twist was predicted \cite{Rybakov2013,Rybakov2016,Leonov2016},
which stabilizes the skyrmionic state in thin films. The proposed
3D texture can be described by the $z$-invariant helical spiral modulated
by an additional azimuthal modulation of the magnetization, $\psi=\sin L_{D}z$,
which resembles to add a Néel type magnetic texture close to the surfaces
\cite{Rybakov2016}. Note, however, that the predicted length of the
chiral surface twist is rather small ($<L_{D}/4$), rendering its
experimental observation challenging. Furthermore, little is known
about additional possible surface modulations, e.g., induced by surface
anisotropies. Another layer of complexity is introduced by surface
modulations occurring during the fabrication of thin magnetic layers.
For instance, surface damage during synthesis or TEM specimen preparation
may lead to a non-magnetic layer.

\section{Experimental\label{sec:Experimental}}

In order to experimentally probe the 3D modulation of magnetic textures
we apply focal series in-line and off-axis EH enabling the reconstruction
of the projections of the lateral components of the magnetic induction
$\int\boldsymbol{B}_{\perp}\mathrm{d}z$, respectively (see the Supplemental
Material, Suppl. I). This implies that cycloidal modulations (and
hence also Néel skyrmions) are invisible in these techniques, if they
are aligned perpendicular to the beam, because the stray fields above
and below the thin film sample cancel the lateral fields within the
sample in projection. Thus, to observe cycloidal modulation the specimen
needs to be tilted with respect to the electron beam.

\begin{figure}
\includegraphics[width=1\columnwidth]{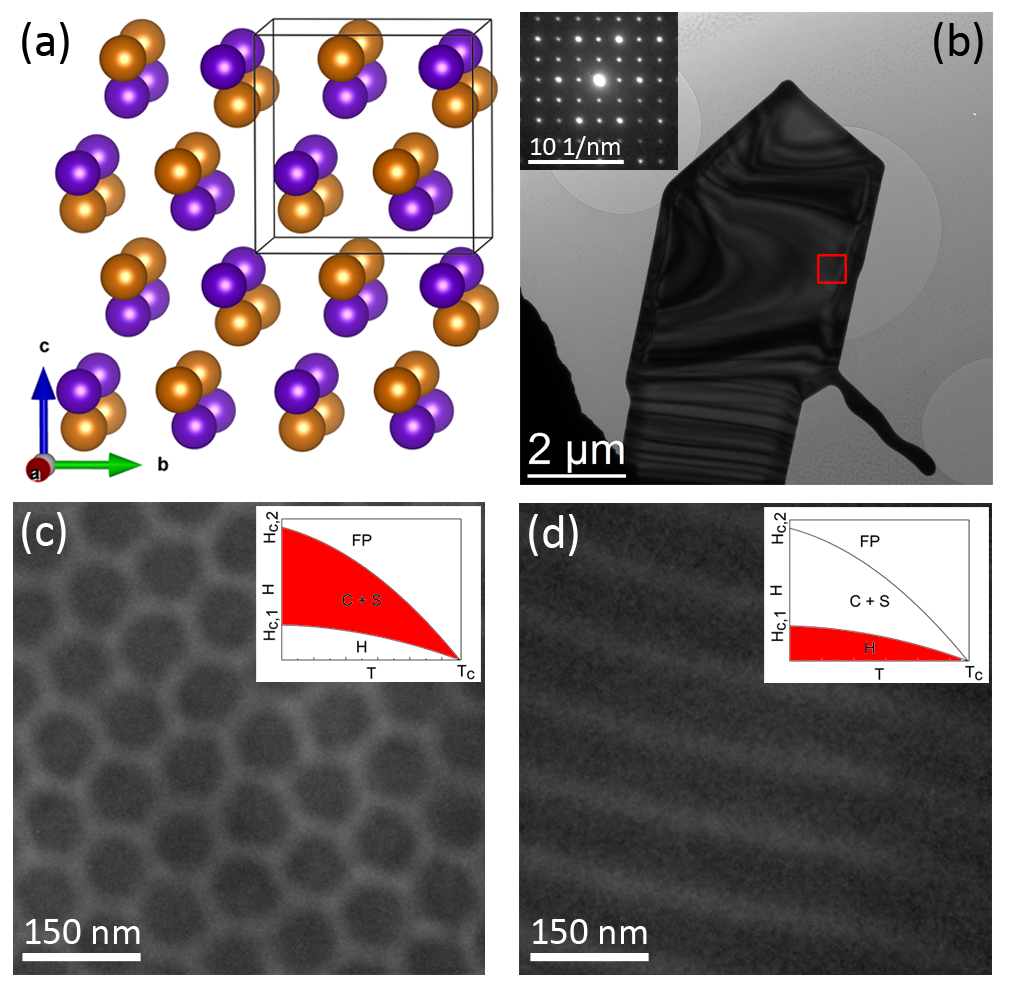}

\caption{(a) Structure of FeGe and $\mathrm{Fe_{0.95}Co_{0.05}Ge}$ in the
cubic B20 phase. Fe and Co atoms are shown in violet and Ge in brown.
(b) TEM image of a $\mathrm{Fe_{0.95}Co_{0.05}Ge}$ nanoplate in $\left[001\right]$
orientation on a holey carbon support with the diffraction pattern
shown in the inset. The red square marks the area investigated by
L-TEM and EH. (c) Skyrmion lattice and (d) helical phase as observed
within the marked area in panel (a) in fields of $43\,mT$ and $0\,mT$,
respectively. The insets show schematically the magnetic phase diagrams
with the corresponding phases marked in red. H, C, S, and FP denote
the helical, cycloidal, skyrmion or field polarized ferromagnetic
phases, respectively. \label{fig: Fig. 2}}
\end{figure}

The skyrmion phase was investigated using a double corrected FEI $\mathrm{Titan^{3}}$
80-300 microscope operated in imaging corrected Lorentz mode (conventional
objective lens turned off) at an acceleration voltage of $300\,kV$.
All measurements were performed at a sample temperature of $90\,K$
using a Gatan double tilt liquid nitrogen cooling holder. Since artifacts
implemented during the sample preparation in the standard FIB preparation
of thin TEM lamellas may alter the magnetic properties of the thin
film, we investigated as-synthesized $\mathrm{Fe_{0.95}Co_{0.05}Ge}$
particles (see the Supplemental Material, Suppl. II). For TEM investigations,
the particles were transferred onto a holey carbon film by swiping
the TEM grid gently over the particles on top of the Ge substrate.
An applied field of $43\,mT$ in out-of-plane direction leads to the
appearance of the skyrmion phase in the slab-like $\mathrm{Fe_{0.95}Co_{0.05}Ge}$
nanoplate (see Fig. \ref{fig: Fig. 2} (d)). A focal series of Lorentz
TEM (L-TEM) images ranging from $+840\,\mu m$ to $-840\,\mu m$ in
focus steps of $84\,\mu m$ of a single isolated nanoplate oriented
along $[001]$ zone axis (see Fig. \ref{fig: Fig. 2} (b)) was recorded.
Reconstruction of the electron wave's phase and hence the magnetic
induction was obtained with the help of a modified Gerchberg-Saxton
type algorithm incorporating affine image registration due to magnification
change and residual shifts as well as rotations building up in such
a long range focal series \cite{Lubk2016}.

To supplement the focal series reconstructions from large field of
views (eventually suffering from a damping of very low and large spatial
frequencies \cite{Lubk2016}), smaller areas of the same nanoplate
were investigated by off-axis EH \cite{Lichte2013}. To this end,
the electron biprism voltage was set to $180\,V$ to produce an overlap
interference width of $500\,nm$ and a holographic interference fringe
spacing of $1.4\,nm$. For hologram recording, an exposure time of
$4\,s$ was employed. Off-axis electron holograms were reconstructed
numerically using a standard Fourier transform based method with sideband
filtering using in-house developed scripts for Gatan Microscopy Suite
(GMS) software package. Contour lines and colour maps were generated
from recorded magnetic phase images to yield magnetic induction maps
(see the Supplemental Material, Suppl.~III). The sample thickness
was determined by means of zero-loss energy-filtered (EF)TEM using
a Gatan Tridiem 865 energy filter yielding a thickness wedge over
the field of view between $100\,nm$ and $150\,nm$ in the case of
in-line EH as well as between $100\,nm$ and $200\,nm$ in the case
of off-axis EH. In the latter case, the thickness measurement was
confirmed by using the phase image (see the Supplemental Material,
Suppl.~IV).

To reveal the $z$-modulation, we recorded a EH tilt series of the
helical phase of the previously investigated nanoplate ranging from
-40\textdegree{} to +30\textdegree{} without magnetic field applied
at both temperatures $90\,K$ and room temperature (see the Supplemental
Material, Suppl.~V). Higher tilt angles could not be attained because
of technical limitations of the liquid nitrogen cooling holder. We
reduced dynamical diffraction contrast by tilting the specimen -7\textdegree{}
out of the $\left[001\right]$ zone axis in direction perpendicular
to the TEM goniometer axis. In order to extract 3D information from
the limited tilt range of projections, we compare the tomographic
data to projected magnetic fields obtained from magnetostatic simulations
(see the Supplemental Material, Suppl.~VII). Hereby, we take into
account several $z$-dependent spin modulations, such as a non-magnetic
surface layer or a chiral surface twist. The $z$-invariant case (i.e.,
surface layer thickness equal zero) is thereby used as reference.

\section{Results}

\begin{figure}[h]
\includegraphics[width=0.9\columnwidth]{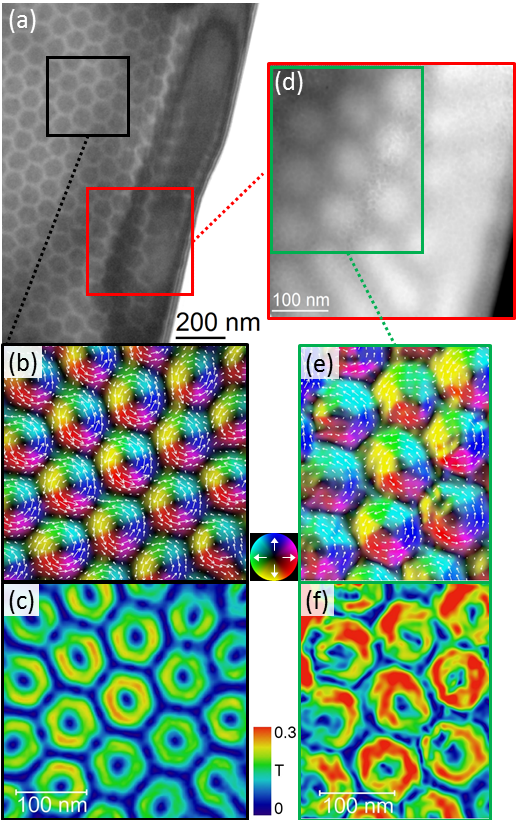}

\caption{\label{fig: Fig. 4}Reconstruction of the projected in-plane component
of the magnetic induction, $\boldsymbol{B}_{\perp}$, within the hexagonal
skyrmion lattice of the $\mathrm{Fe_{0.95}Co_{0.05}Ge}$ nanoplate
using in-line (a-c) and off-axis holography (d-f). (a) L-TEM image
at $-84\,\mathrm{\mu m}$ defocus showing the skyrmions as dark contrast.
(d) Phase image at the position indicated by the red square in (a).
(b,e) Mapping of the direction of $\boldsymbol{B}_{\perp}$ by combining
a vector plot (white arrows) and a false color image. (c,f) False
colour mapping of the magnitude of $\boldsymbol{B}_{\perp}$.}
\end{figure}

\begin{figure}
\includegraphics[width=1\columnwidth]{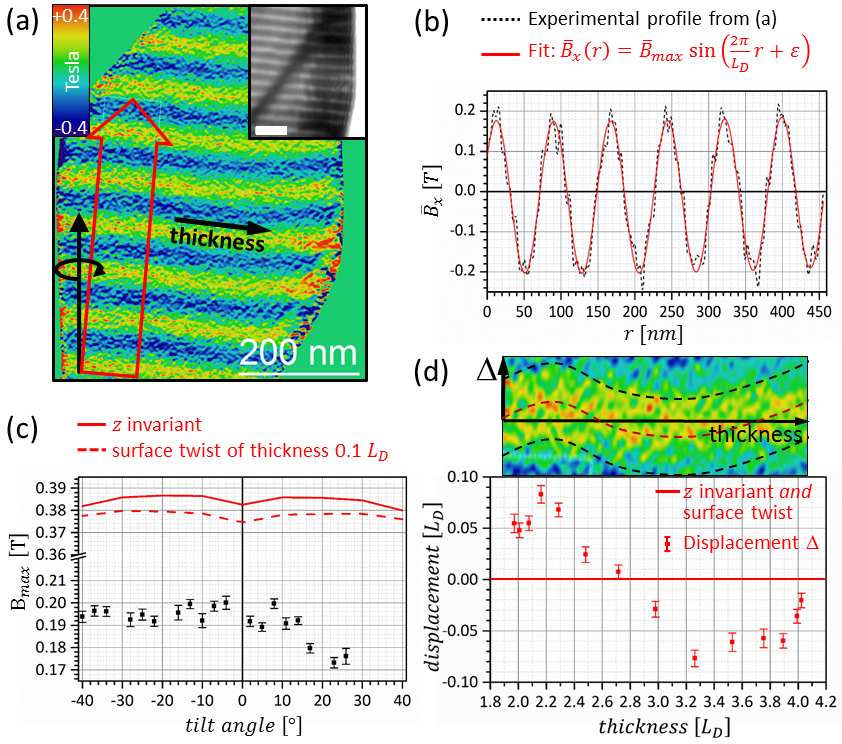}

\caption{Dependence of the in-plane magnetic induction on the specimen tilt
around the propagation axis of the helical phase in a $\mathrm{Fe_{0.95}Co_{0.05}Ge}$
nanoplatelet. (a) In-plane component $B_{x}$ (as determined from
off-axis holography) normalized to the specimen thickness. The tomographic
tilt axis and the direction of the thickness gradient are indicated
by black arrows. The inset shows the corresponding L-TEM image (scale
bar = $200\,nm$). (b) Line profile of the projected in-plane magnetic
induction $\bar{B}_{x}$ along the red arrow (and averaged along its
width) in (a) and sinusoidal fit using the function denoted above.
(c) $\boldsymbol{B}$-field amplitudes $\bar{B}_{max}$ obtained from
likewise determined fits to the induction maps as function of the
tilt angle. Simulations for ($z$-invariant) pure helical spirals
(solid line) and with chiral surface twist (dashed line) are included
for comparison. (d) Amplitude $\Delta$ of the undulation of the magnetic
stripe contrast (perpendicular to the helical axis) around its mean
value as function of the thickness in units of $L_{D}$. The anisotropically
magnified stripe image shown as an example above indicates how $\Delta$
is determined. The error bars in (c) and (d) are given by the fit
error. \label{fig: Fig. 3}}
\end{figure}

Fig.~\ref{fig: Fig. 4}~(a) depicts a L-TEM micrograph in underfocus
showing the hexagonal skyrmion lattice as dark contrast, which is
one image of the focal series used for in-line holography reconstruction
of the object exit wave in amplitude and phase (see Sect.~\ref{sec:Experimental}
for acquisition and reconstruction parameters). The projected in-plane
components of magnetic induction ($\boldsymbol{B}_{\mathbf{\perp}}^{\mathrm{proj}}\left(x,y\right)$)
were computed from the spatial derivative of the reconstructed phase
image (Supplemental Material, Eq. (2)). The knowledge of the projected
thickness, which we determined by zero-filtered EFTEM, enables us
to compute the $\boldsymbol{B}_{\perp}\left(x,y\right)$ components
averaged along $z$-direction, i.e., $\boldsymbol{\overline{B}}_{\perp}\left(x,y\right)$.
Figs.~\ref{fig: Fig. 4}~(b,c) show magnetic induction maps $\boldsymbol{\overline{B}}_{\perp}\left(x,y\right)$
in cylindrical coordinate representation visualizing the spin texture
of the skyrmions by \textbf{$\overline{B}_{\phi}\left(x,y\right)$}
(Fig.~\ref{fig: Fig. 4}~(b)) and their donut-shaped magnitude by
\textbf{$\overline{B}_{r}\left(x,y\right)$}. Likewise, we observed
magnetic induction maps (Figs.~\ref{fig: Fig. 4}~(e,f)) from a
phase image reconstructed by off-axis EH (Fig.~\ref{fig: Fig. 4}~(d))
on the same $\mathrm{Fe_{0.95}Co_{0.05}Ge}$ nanoplate. Comparing
the results of the two holographic methods, we measure a slightly
higher magnetic induction \textbf{$\overline{B}_{r}\left(x,y\right)$}
with a slightly higher spatial resolution in the case of off-axis
holography. However, we consistently observe a reduction of the $\boldsymbol{B}$-fields
($\overline{B}_{max}=\left(0.2\,...\,0.3\right)\,T$) with respect
to the $z$-invariant case ($\overline{B}_{max}=0.43\,T$) obtained
from magnetostatic simulations. Note that we observe similar reductions
in a variety of FeGe samples subject to different preparation histories
(e.g., as-synthesized nanocrystals, FIB lamellas). Consequently, we
can exclude the presence of a non-magnetic surface layer of approximately
$40\,nm$ as the principle reduction mechanism. In order to clarify
the origin of the reduced values a 3D reconstruction of the magnetic
induction by means of electron holographic tomography \cite{Wolf_CoM27(2015)6771}
would be required.
In case of the skyrmionic lattice a tomographic investigation of the
3D modulation is currently experimentally unfeasible, because this
necessitates an externally applied out-of-plane magnetic field to
be tilted with the sample. In the current experimental setup, the
skyrmions align along the magnetic field of the objective lens which
has a fixed orientation along the optical axis. In order to overcome
these pertaining experimental challenges in-situ magnetic vector field
application devices and auxiliary magnetic signals such as EMCD \cite{Schattschneider2006,Schneider2016,Pohl2017,Edstrom2016}
would be helpful. In the following, we therefore resort to acquiring
a tilt series of the helical phase stabilizing without applied external
field. A representative holographic $\boldsymbol{B}$-field reconstruction
of the helical phase is depicted in Fig.~\ref{fig: Fig. 3}~(a,b).
Accordingly, we observe sinusoidal modulations of the projected lateral
$\boldsymbol{B}$-field component with a period of $77\,nm$ corresponding
to spiral magnetic textures aligned in plane (see the Supplemental
Material, Suppl.~VI). Unfortunately, upon tilting rather large local
variations occur in the phase images (e.g., bending fringes visible
in the inset of Fig~\ref{fig: Fig. 3}~(a)) that are related to
changes in the dynamic scattering conditions of the $175\,nm$ thick
crystalline sample in the region of interest (ROI). Likewise, the
phase images taken at room temperature required to determine the thickness
maps for each tilt, suffer from dynamical diffraction contrast and
need to be treated with care (see the Supplemental Material, Suppl.~V).
After normalizing the projected magnetic fields with corresponding
thickness maps, we consistently observe a reduction of the $\boldsymbol{B}$-fields
($\overline{B}_{\mathrm{max}}=0.2\,T$) with respect to the $z$-invariant
case ($\overline{B}_{\mathrm{max}}=0.43\,T$) as determined in the
case of the skyrmion texture. Consequently, a similar 3D modulation
as for the skyrmion lattice is expected under field-free conditions.

To gain more insight in the underlying spin texture, we analyze the
magnetic contrast as function of the tilt angles (Fig.~\ref{fig: Fig. 3}~(c)).
In the investigated nanoplatelet, the helices are aligned almost parallel
to the tilt axis (mistilt of 5\textdegree{} as depicted in Fig.~\ref{fig: Fig. 3}
(a)). This orientation is well-suited to identify any 3D spin modulation,
such as the previously discussed mixing of helical (i.e., Bloch type)
and cycloic (Néel type) spiral spin textures in the chiral surface
twist, as additional contrast modulation of the stripe pattern in
the tilt series. For the contrast measurement, we shifted the ROI
in each phase image of the tilt series such that the same mean thickness
of $175\,nm$ was achieved. This suppresses the possible influence
of the thickness on the projected fields. The tilt series show a nearly
constant value of $\overline{B}_{\mathrm{max}}$ for tilt angles from
-40\textdegree{} to 12\textdegree{} and a strong drop by more than
$10\,\%$ at higher tilt angles. Two different scenarios have been
evaluated to clarify the 3D spin configuration: (i) in case of a pure
helical spiral without surface twist (solid red line) and upon tilting,
the thickness-normalized fringe contrast is almost constant except
for a slight modulation around zero tilt due to the above-mentioned
mistilt of -7\textdegree . This also causes a reduction of $\overline{B}_{\mathrm{max}}$
from $0.43\,T$ to $0.38\,T$. (ii) Cycloic-like modulations in a
surface twist layer with a thickness of $0.1\,L_{D}$ lead mainly
to an additional contrast damping (dashed red line) of approximately
$3\,\%$ percent, which is, however, significantly smaller than experimentally
observed (Fig.~\ref{fig: Fig. 3}~(c)). The measured $50\,\%$ reduction
of the projected fields may only be explained by additional $z$-dependent
modulations, spanning larger sections of the film. Evaluating the
local fringe position (lateral phase) in a representative induction
map from the tilt series (cf. Fig.~\ref{fig: Fig. 3}~(a)) and correlating
the latter with the corresponding thickness map, we also observed
lateral in-plane) displacements of the helical stripes as a function
of the overall thickness (see Fig.~\ref{fig: Fig. 3}~(d) and scheme
above). Such an undulation points to a lateral shift of the helix
as a function of the $z$-coordinate, which in turn would on the one
hand lead to an additional contrast damping, while on the other hand,
would explain the observed asymmetric dependence of the contrast on
the tilt angle (Fig.~\ref{fig: Fig. 3}~(c)). Such surface-related
modulations of the spin texture may be stabilized by additional (surface)
anisotropies in the above free energy functional.

In summary, we carried out a quantitative electron holographic reconstruction
of the projected in-plance magnetic induction in $\mathrm{Fe_{0.95}Co_{0.05}Ge}$
examined under various tilt directions. We show that these projected
magnetic fields are significantly smaller than the fields expected
for both $z$-invariant Bloch Skyrmions and the theoretic predictions
of chiral surface twists in the thin surface layers of such helimagnets.
Although this finding cannot be accounted for by any present model
of spin structures, it clearly shows that the underlying magnetic
structure substantially deviates from that of a regular Bloch skyrmion
in major sections of the film in $z$-direction. Analyzing the thickness
dependence of the projeted in-plane field pattern of the helical phase
further reveals modulations of this structure also in the $x$-$y$-plane.
Hence, rather than supporting the model of a homogeneous skyrmion
lattice, the results of the present investigations can only be understood
by assuming modulations of the skyrmionic structure in all three dimensions
throughout the helimagnetic film.

\begin{acknowledgments}
We thank T. Gemming for providing to us
the Gatan double tilt liquid nitrogen cooling holder. The authors
are indebted to A. Pöhl and T. Sturm for the preparation of the TEM
samples. We thank U. Rößler, A. Leonov and A. Bogdanov for fruitful
discussions. This research is supported by NSF grant ECCS-1609585.
MJS also acknowledges support from the NSF Graduate Research Fellowship
Program grant number DGE-1256259. AL and DW have received funding
from the European Research Council (ERC) under the Horizon 2020 research
and innovation programme of the European Union (grant agreement No
715620).
\end{acknowledgments}

\bibliographystyle{apsrev4-1}

%
\end{document}


\title{Supplementary Information: Induction Mapping of the 3D-Modulated
Spin Texture of Skyrmions in Thin Helimagnets}

\author{S. Schneider}

\affiliation{Institute for Metallic Materials, IFW Dresden, Helmholtzstr. 20,
01069 Dresden, Germany}

\affiliation{Institut für Festkörper- und Materialphysik, Technische Universität
Dresden, 01062 Dresden, Germany}

\author{D. Wolf}

\affiliation{Institute for Solid State Research, IFW Dresden, Helmholtzstr. 20,
01069 Dresden, Germany}

\author{M. J. Stolt}

\affiliation{Department of Chemistry, University of Wisconsin-Madison, 1101 University
Avenue, Madison, Wisconsin 53706, United States}

\author{S. Jin}

\affiliation{Department of Chemistry, University of Wisconsin-Madison, 1101 University
Avenue, Madison, Wisconsin 53706, United States}

\author{D. Pohl}

\affiliation{Institute for Metallic Materials, IFW Dresden, Helmholtzstr. 20,
01069 Dresden, Germany}

\affiliation{Dresden Center for Nanoanalysis, Technische Universität Dresden,
01062 Dresden, Germany}

\author{B. Rellinghaus}

\affiliation{Institute for Metallic Materials, IFW Dresden, Helmholtzstr. 20,
01069 Dresden, Germany}

\affiliation{Dresden Center for Nanoanalysis, Technische Universität Dresden,
01062 Dresden, Germany}

\author{M. Schmidt }

\affiliation{Department Chemical Metal Science, Max Planck Institute for Chemical
Physics of Solids, Nöthnitzer Str. 40, 01187 Dresden, Germany}

\author{B. Büchner}

\affiliation{Institute for Solid State Research, IFW Dresden, Helmholtzstr. 20,
01069 Dresden, Germany}

\author{S. T. B. Goennenwein}

\affiliation{Institut für Festkörper- und Materialphysik, Technische Universität
Dresden, 01062 Dresden, Germany}

\affiliation{Center for Transport and Devices of Emergent Materials, Technische
Universität Dresden, 01062 Dresden, Germany}

\author{K. Nielsch}

\affiliation{Institute for Metallic Materials, IFW Dresden, Helmholtzstr. 20,
01069 Dresden, Germany}

\affiliation{Institute of Materials Science, Technische Universität Dresden, Helmholtzstr. 7,
01069 Dresden, Germany}

\author{A. Lubk}

\affiliation{Institute for Solid State Research, IFW Dresden, Helmholtzstr. 20,
01069 Dresden, Germany}
\maketitle

\section{Magnetic Holography}

In off-axis electron holography the magnetic phase shift between the
electron wave passing the object containing the skyrmions and the
vacuum reference wave is given by the Aharonov-Bohm effect, i.e.,
by the magnetic flux enclosed between superimposing paths of the object
and reference wave (e.g., \cite{Lichte2013}). Consequently, the spatial
derivative of the reconstructed phase reads
\begin{eqnarray}
\partial_{x,y}\varphi_{\mathrm{mag}}\left(\boldsymbol{r}_{\bot}\right) & = & \frac{e}{\hbar}\partial_{x,y}\oint\boldsymbol{A}\left(\boldsymbol{r}\right)\mathrm{d}\mathbf{s}\\
 & = & \frac{e}{\hbar}\partial_{x,y}\oiint\iint\boldsymbol{B}\left(\boldsymbol{r}\right)\mathrm{d}\mathbf{S}\nonumber \\
 & = & \frac{e}{\hbar}\int_{-\infty}^{\infty}\begin{pmatrix}B_{y}\left(\boldsymbol{r}_{\bot}\right)\\
-B_{x}\left(\boldsymbol{r}_{\bot}\right)
\end{pmatrix}\mathrm{d}z\,,\label{eq:PhaseGrad_vs_B}
\end{eqnarray}
facilitating the computation of the projected magnetic fields. The
last line of the above relation also holds in the case of Inline holography.
To separate electric and magnetic phase shifts two images with flipped
specimen are recorded \cite{Kasama2011}. Subtracting both allows
to remove the electric phase shift, which is not changing sign under
spatial inversion (as opposed to the magnetic one, which changes sign).

\section{Sample Preparation}

Single crystals of FeGe were grown by chemical transport reaction.
The transport experiment was carried out from the elements (Fe $99.998\,\%$
Alfa Aesar, Ge $99.9999\,\%$ Alfa Aesar) in an evacuated quarz tube
in a temperature gradient from $580\,{^\circ}C$ (source) to $540\,{^\circ}C$
(sink). Iodine was used as transport agent. Selected crystals were
characterized by EDXS, X-ray powder, and X-ray single crystal diffraction.

$\mathrm{Fe_{0.95}Co_{0.05}Ge}$ particles were fabricated by chemical
vapor deposition process analogous to the one described in \cite{Stolt2017}
used to form B20 FeGe nanowires. This CVD process, which was carried
out at $550\,{^\circ}C$, was adjusted slightly by the addition of
$\backsim100$ mg of $\mathrm{CoI_{2}}$ to the alumina source boat.
This addition of $\mathrm{CoI_{2}}$ resulted in both the small introduction
of Co (<10\%) into the final particles as well as the drastic change
in morphology from nanowires to the slab-like particle shape.

\section{Off-Axis Holography Reconstruction}

\begin{figure}[H]
\begin{centering}
\includegraphics[width=0.8\columnwidth]{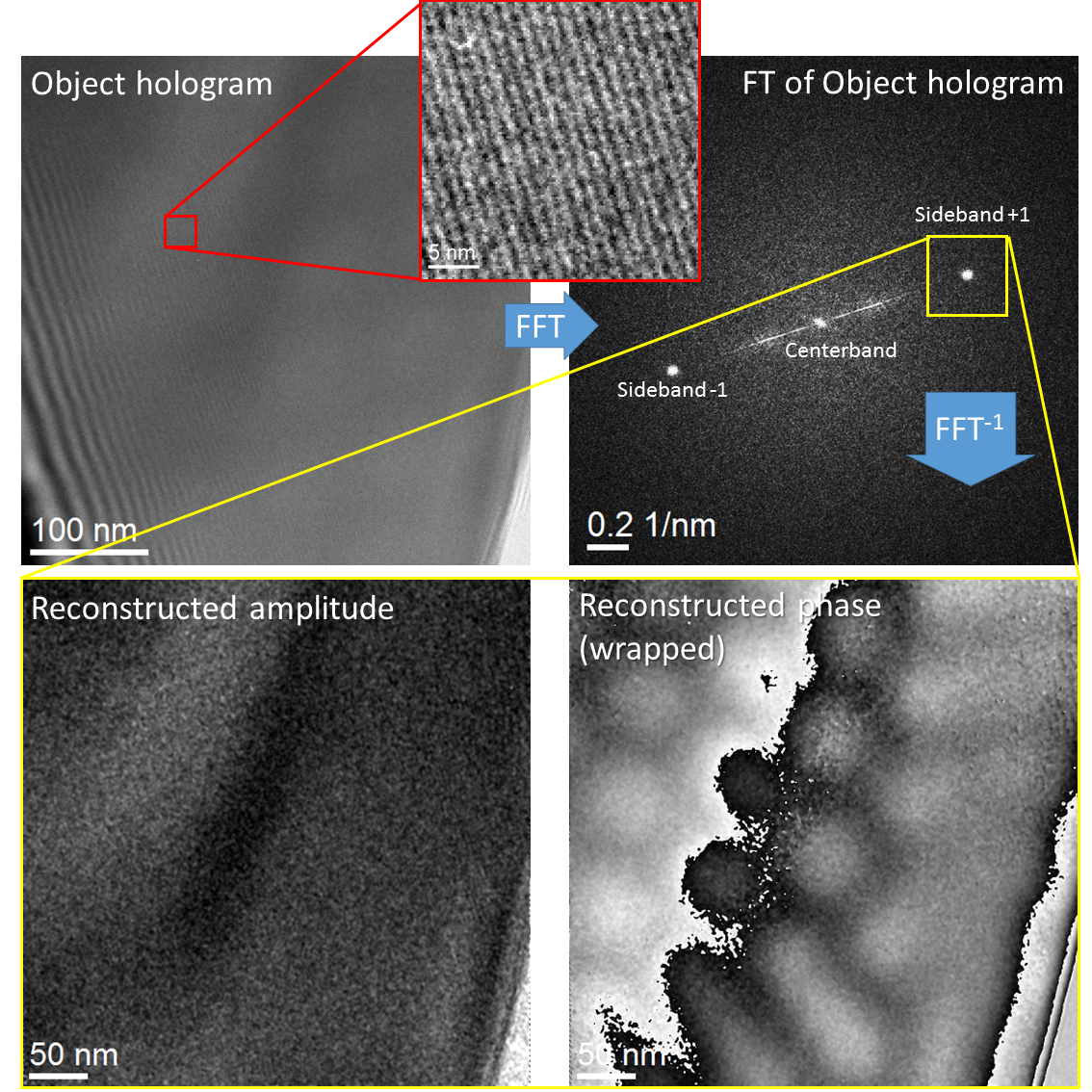}
\par\end{centering}
\caption{\label{fig:Object-hologram}Object hologram reconstruction by Fourier
method yielding the object exit wave, i.e., amplitude and phase image.
Specimen: edge of a $\mathrm{Fe_{0.95}Co_{0.05}Ge}$ particle.}
\end{figure}

Fig. \ref{fig:Object-hologram} depicts the object hologram, from
which the object exit wave, amplitude and phase, is reconstructed.
The inset shows the hologram fringes within the $\mathrm{Fe_{0.95}Co_{0.05}Ge}$
specimen with reasonable contrast of $5.5\%$ at $1.4\,nm$ fringe
spacing ($6.6\,\nicefrac{hologram\,pixels}{fringe}$). The Fourier-based
reconstruction envolves fast Fourier transformation (FFT) of the hologram,
centering and cutting out one sideband with a numerical mask (here:
sinc-modulated square mask with $0.25\,nm^{-1}$ edge length) and
inverse FFT of the cut sideband. Furthermore, the phase image had
to be unwrapped further to obtain the full phase shift larger than
$2\pi$, and the phase offset was set to zero in vacuum (lower right
edge of the image).

\section{Thickness Determination}

\begin{figure}[H]
\centering{}\includegraphics[width=1\columnwidth]{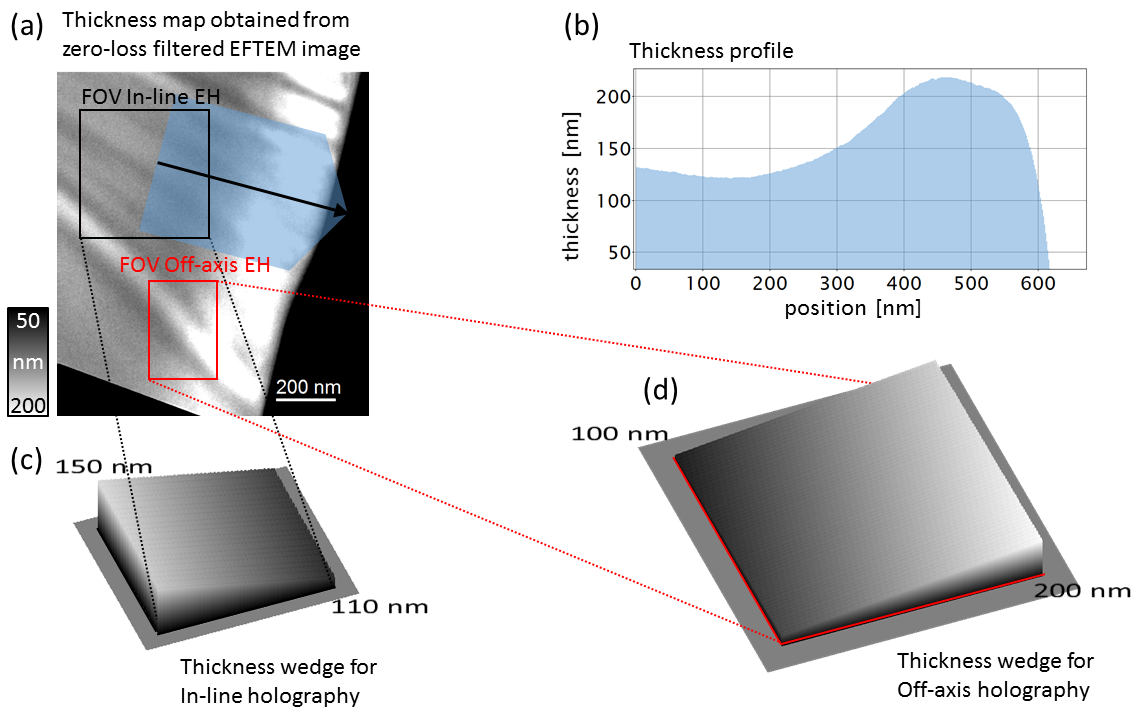}\caption{\label{fig:Thickness-measurement}Thickness measurement from zero-loss
filtered EFTEM image of the $\mathrm{Fe_{0.95}Co_{0.05}Ge}$ particle
which is investigated by electron holography. (a) Thickness map computed
with Eq. \ref{eq:tol}. (b) Thickness profile towards the edge according
to the arrow marked in (a). (c) Thickness wedge obtained by linear
regression within the field of view (FOV) for in-line holography studies.
(d) Thickness wedge obtained by linear regression within the field
of view (FOV) for off-axis holography studies.}
\end{figure}

The object thickness was measured by evaluating zero-loss TEM micrographs
with 10~eV energy slit width denoted here as $I_{\mathrm{in}}$.
In addition, unfiltered images were acquired to take into account
the contribution stemming from elastical scattering denoted here as
$I_{\mathrm{el}}$. The thickness $t$ over inelastic mean-free-path
$\lambda_{\mathrm{in}}$ map can be computed using the relation

\begin{equation}
\frac{t\left(x,y\right)}{\lambda_{\mathrm{in}}}=-\ln\left(\frac{I_{\mathrm{in}}\left(x,y\right)}{I_{\mathrm{el}}\left(x,y\right)}\right)\label{eq:tol}
\end{equation}
With the knowledge of $\lambda_{\mathrm{in}}=140\,nm$ which we computed
from the laws of plasmon scattering \cite{Egerton1996}, we are able
to determine a thickness map $t\left(x,y\right)$ as depicted in Fig.
\ref{fig:Thickness-measurement}

\begin{figure}[h]
\begin{centering}
\includegraphics[width=1\columnwidth]{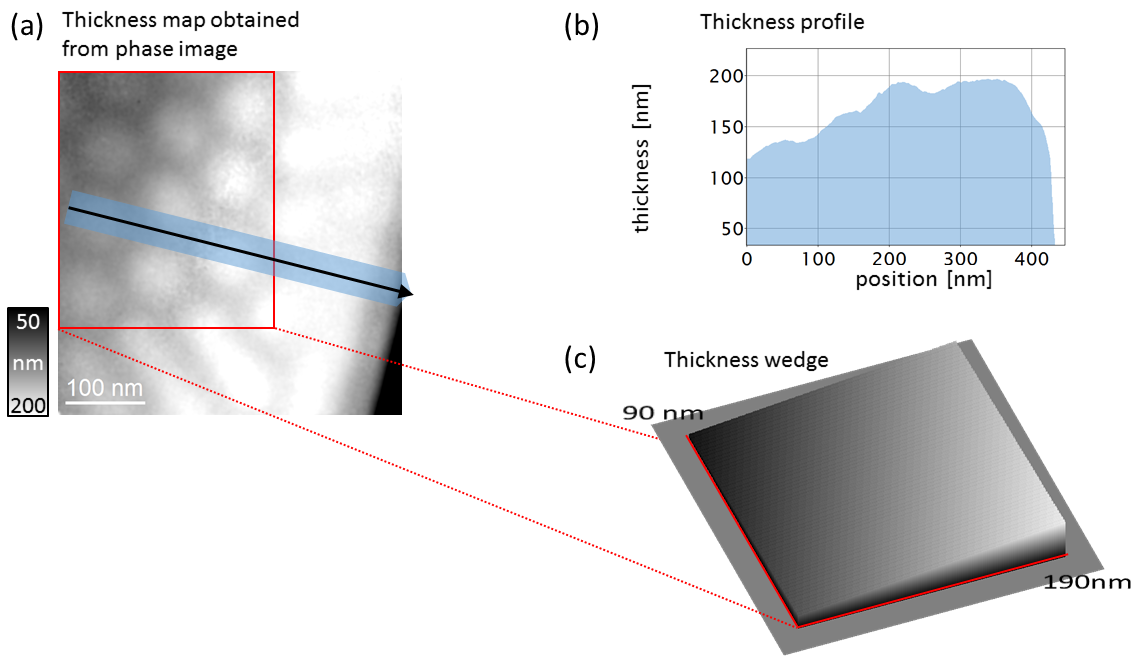}
\par\end{centering}
\caption{\label{fig:Thickness-measurement-from-phase}Thickness measurement
from phase image of the $\mathrm{Fe_{0.95}Co_{0.05}Ge}$ platelet
which is investigated by electron holography. (a) Thickness map computed
with Eq. \ref{eq:phase}. (b) Thickness profile towards the edge according
to the arrow marked in (a). (c) Thickness wedge obtained by linear
regression within the field of view (FOV) for off-axis holography
studies.}
\end{figure}

Likewise, the specimen thickness was determined exploiting the relation
between the phase shift $\varphi\left(x,y\right)$ and the thickness
distribution $t\left(x,y\right)$, \textit{i.e.},

\begin{equation}
\varphi\left(x,y\right)=C_{E}V_{0}t\left(x,y\right)\label{eq:phase}
\end{equation}
with the interaction constant $C_{E}=0.0065\,\nicefrac{rad}{Vnm}$
at $300\,kV$ accelaration voltage and the mean inner potential (MIP)
$V_{0}$. The latter was determined to $V_{0}=13.4\,V$ using, (i)
isolated atomic potentials for Fe and Ge (parametrization of Weickenmeyer
and Kohl \cite{Weickenmeier1991}), (ii) a lattice constant of the
cubic unit cell of $4.7\,\mathring{\mathrm{A}}$, and (iii) two atoms
per unit cell for Fe and Ge, respectively. The result shown in Fig.
\ref{fig:Thickness-measurement-from-phase} reveals that the obtained
thickness wedge (\ref{fig:Thickness-measurement-from-phase}(c)) deviates
only by about $10\,nm$ compared to the corresponding one obtained
by EFTEM (Fig. \ref{fig:Thickness-measurement}(d)).

\section{Holographic Tilt Series}

\begin{figure}[H]
\begin{centering}
\includegraphics[width=0.9\columnwidth]{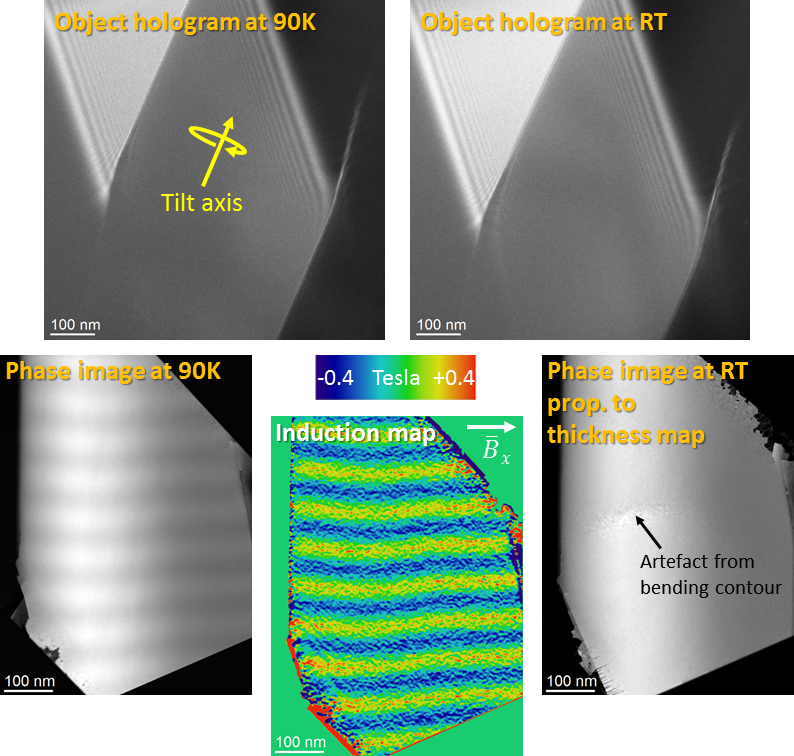}
\par\end{centering}
\caption{\label{fig:Holos etc}Holograms, reconstructed phase images, and obtained
horizontal $\boldsymbol{B}$-field component normalized thickness
map of the $\mathrm{Fe_{0.95}Co_{0.05}Ge}$ platelet at a tilt angle
of $-27\text{\textdegree}$.}
\end{figure}

In order to reveal a potential out-of-plane $\boldsymbol{B}$-field
component of the helical phase in the $\mathrm{Fe_{0.95}Co_{0.05}Ge}$
platelet, we investigated magnetic phase images recorded under different
specimen tilt angles. The orientation of the tilt axis was chosen
parallel to the propagation axis of the helical phase. We performed
the following workflow:
\begin{enumerate}
\item Acquisition of hologram tilt series from -30\textdegree{} to +39\textdegree{}
in 3\textdegree{} steps at room temperature (RT)
\item Acquisition of hologram tilt series from -30\textdegree{} to +39\textdegree{}
in 3\textdegree{} steps at 90 K
\item Reconstruction of phase images from holograms
\item Phase unwrapping and setting phase offset in the vacuum region of
the phase image to 0 rad
\item Displacements correction of phase images within the tilt series, as
well as between phase images at RT and 90 K
\item Calculation of derivatives of phase images in parallel ($y$-) direction
to the tilt axis to obtain the projected $B$-component perpendicular
to the tilt axis ($B_{x}^{\mathrm{proj}}$)
\item Calculation of thickness map from phase images at RT as explained
in the previous section
\item Calculation of averaged $B_{x}$ by
\[
\bar{B}_{x}\left(x,y\right)=\frac{B_{x}^{\mathrm{proj}}\left(x,y\right)}{t\left(x,y\right)}
\]
\end{enumerate}
In Fig. \ref{fig:Holos etc} holograms, reconstructed phase images,
and the horizontal $\boldsymbol{B}$-field component obtained by the
above-mentioned workflow for a certain specimen tilt angle are depicted.

\begin{figure}
\begin{centering}
\includegraphics[width=0.9\columnwidth]{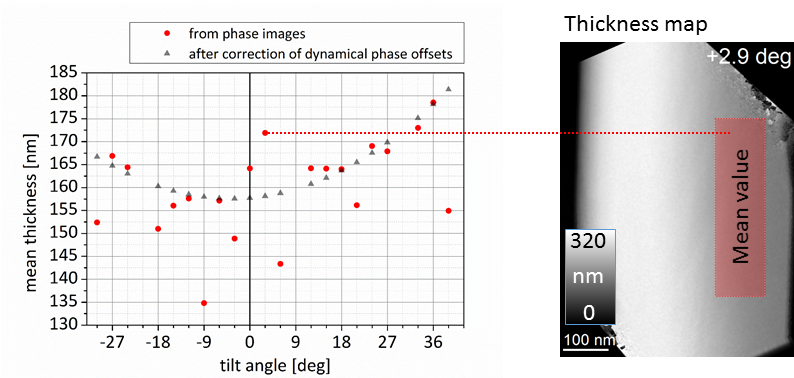}
\par\end{centering}
\caption{\label{fig:Thickness}Projected thickness as a function of projection
angle (sample orientation). The mean thickness was obtained in the
region of the thickness map indicated by the red rectangle.}
\end{figure}

We encountered the problem that the phase images of the tilt series
suffered from diffraction contrast, which lead to significant deviations
in the phase signal and the determined thickness maps. Moreover, since
bending contours caused by dynamical diffraction disturbed the phase
images locally, the corresponding regions had to be excluded from
the analysis. We tackled the first problem by plotting the projected
thickness averaged over corresponding regions as a function of projection
tilt angle, thereby identifying the ``actual'' trend of the projected
thickness by a quadratic fit (Fig. \ref{fig:Thickness}). Note the
negative systematic underestimation of the thickness at larger tilt
angles.

\section{Magnetic induction maps}

\begin{figure}
\includegraphics[width=1\columnwidth]{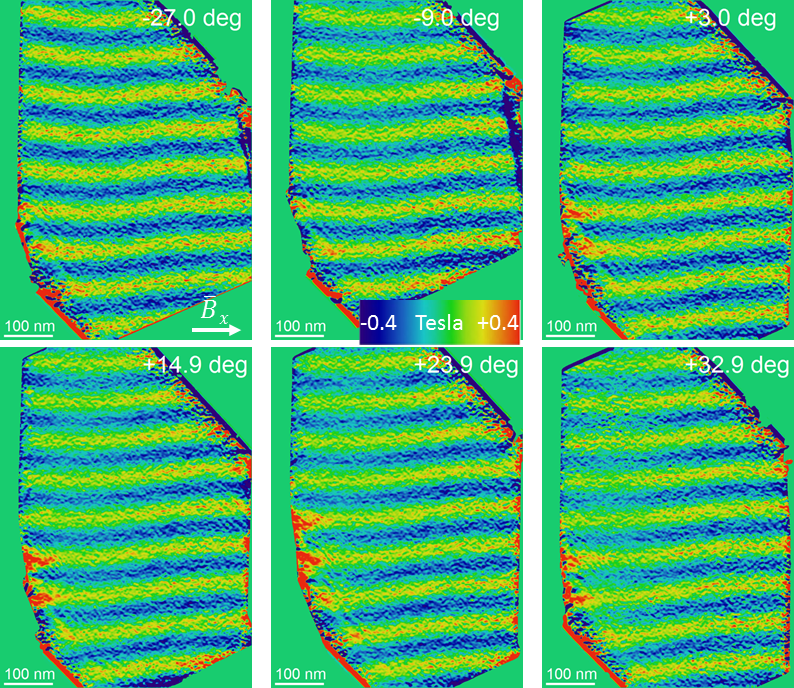}

\caption{\label{fig:Induction-maps vs tilt}Induction maps of the $\mathrm{Fe_{0.95}Co_{0.05}Ge}$
nanoplate in the helical phase taken at different tilt angles.}
\end{figure}

Fig. \ref{fig:Induction-maps vs tilt} shows representative magnetic
induction maps of the tilt series after normalization with the corresponding
thickness maps. It demonstrates that the helical phase modulations
are not straight but they are slightly bended in lateral direction.
Since the thickness increases significantly towards the object edge
on the left hand side (compare Fig. \ref{fig:Thickness} right), a
possible explanation of this bending could be an underlying modulation
in depth $(z)$ direction.

\section{Magnetostatics}

The saturation magnetization of a $\mathrm{Fe_{0.95}Co_{0.05}Ge}$
powder measured by SQUID amounts to
\begin{equation}
M_{s}=3.45\times10^{5}\,\frac{{A}}{m}\,.
\end{equation}
From that we may derive the maximal magnetic induction expected in
the $z$-invariant spin modulation by taking into account that Bloch
like spin modulations are magnetic charge free. With no in-plane applied
field this yields

\begin{eqnarray}
B_{\bot,\mathrm{max}} & = & \mu_{0}(M+H) \\ 
&\stackrel{H=0}{=} & 4\pi\times10^{-7}\,\frac{{Vs}}{Am}\,\mathrm{\cdot\,}3.45\times10^{5}\,\frac{{A}}{m} =0.43\,T.\nonumber
\end{eqnarray}
For the quantitative simulation of the complete magnetic field distribution
(and hence the projected fields) in and around a thin slab of FeGe
carrying a 3D modulated helical or Bloch Skyrmion texture we solved
the well-known system of magnetostatic equations

\begin{align}
\nabla\boldsymbol{M} & =\triangle\Phi_{M}\\
\boldsymbol{H} & =\nabla\Phi_{M}\nonumber \\
\boldsymbol{B} & =\mu_{0}\left(\boldsymbol{H}+\boldsymbol{M}\right)\nonumber
\end{align}
with a scalar potential $\Phi_{M}$ numerically. More specifically,
we started out by setting up a 3D magnetic texture $\boldsymbol{M}$
including various surface modulations, i.e., chiral surface twist,
and $z$-dependent shifts. Subsequently, we computed the scalar magnetic
potential $\Phi_{M}$, the magnetic field strength $\boldsymbol{H}$
and finally the magnetic induction $\boldsymbol{B}$. The differential
equations involved are solved in Fourier space, which implies that
periodic boundary conditions have been assumed. In order to suppress
artefacts due the this choice of boundary conditions, our computations
have been carried out in a sufficiently large supercell containing
the slab. To compare the simulation with the experimentally reconstructed
projections of the $\boldsymbol{B}$-field under different tilt angles,
we numerically integrated the magnetic induction into the corresponding
directions.

\bibliographystyle{unsrtnat}